\pdfoutput=1
 \documentclass[prd,twocolumn,nofootinbib,superscriptaddress,preprintnumbers,amsmath,amssymb]{revtex4-1}
\allowdisplaybreaks[4]

\usepackage{ascmac}
\usepackage[breaklinks=true]{hyperref}
\usepackage{mathtools}

\usepackage{graphicx}
\usepackage{xcolor}

\usepackage{amsmath,amssymb,amsthm}
\usepackage{amscd} 
\usepackage{mathrsfs} 

\DeclareMathOperator{\diag}{diag}

\DeclareMathOperator{\rank}{rank}

\DeclareMathOperator{\vol}{vol}

\newcommand{\rme}{\mathrm{e}}
\newcommand{\rmd}{\mathrm{d}}
\newcommand{\rmi}{\mathrm{i}}

\newcommand{\corr}{\leftrightarrow}

\newcommand{\vevs}[1]{\langle #1 \rangle}

\newcommand{\er}[1]{Eq.~\eqref{#1}}
\newcommand{\ers}[1]{Eqs.~\eqref{#1}}

\newcommand{\bb}{\mathbb}

\renewcommand{\b}{\bar}

\newcommand{\bs}{\boldsymbol}

\newcommand{\fr}{\frac}

\newcommand{\der}{\partial}

\newcommand{\wed}{\wedge}
\newcommand{\bmx}{\left(\begin{matrix}}
\newcommand{\emx}{\end{matrix}\right)}
\newcommand{\mtx}[1]{\bmx #1 \emx}

\newcommand{\weds}{\wedge \cdots \wedge}
\newcommand{\sectaps}{\section}

\begin{document}
\preprint{KEK-TH-2398}
\title{Unstable Nambu-Goldstone modes}

\author{Naoki Yamamoto}
\email{nyama@rk.phys.keio.ac.jp}
\affiliation{Department of Physics, Keio University, 
Yokohama 223-8522, Japan}

\author{Ryo Yokokura}
\email{ryokokur@post.kek.jp}
\affiliation{KEK Theory Center,
Tsukuba 305-0801, Japan}
\affiliation{Research and Education Center for
Natural Sciences, Keio University, 
Yokohama 223-8521, Japan}

\date{\today}

\begin{abstract}
Nambu-Goldstone (NG) modes for 0-form and higher-form symmetries can become unstable in the presence of background fields. Examples include the instability of a photon with a time-dependent axion background or with a chirality imbalance, known as the chiral plasma instability, and the instability of a dynamical axion with a background electric field. 
We show that all these phenomena can be universally described by a symmetry algebra for 0-form and higher-form symmetries.
We prove a counting rule for the number of unstable NG modes in terms of correlation functions of broken symmetry generators.
Based on our unified description, we further give a simple new example where one of the NG modes associated with the spontaneous 0-form symmetry breaking $U(1) \times U(1) \to \{1\}$ becomes unstable.

\end{abstract}

\maketitle
 \setcounter{footnote}{0}%
\renewcommand{\thefootnote}{$*$\arabic{footnote}}%

\section{Introduction}%
Spontaneous symmetry breaking is one of the important notions 
in modern physics. 
In particular, spontaneous breaking of continuous global symmetries leads to Nambu-Goldstone (NG) modes~\cite{Nambu:1961tp,Goldstone:1961eq,Goldstone:1962es}, which are gapless excitations that dominate low-energy physics.
Various gapless excitations can be identified as NG modes, such as phonons in superfluids, pions in hadron physics, axions or axionlike particles in particle physics and condensed matter physics, and so on.
While the counting rule for the number of NG modes was originally formulated in relativistic systems possessing Lorentz invariance~\cite{Nambu:1961tp,Goldstone:1961eq,Goldstone:1962es}, 
Lorentz invariance is usually absent in realistic materials.
The extension to nonrelativistic systems without Lorentz invariance~\cite{Nielsen:1975hm, Miransky:2001tw, Schafer:2001bq, Nambu:2004yia, Watanabe:2011ec} was also established~\cite{Watanabe:2012hr, Hidaka:2012ym, Watanabe:2014fva}.
There, the number of NG modes, which may be reduced in the absence of Lorentz invariance, can be counted by using symmetry algebra.

Recently, the notion of symmetries has been extended to higher-form symmetries~\cite{Gaiotto:2014kfa} 
(see also Refs.~\cite{Batista:2004sc,Pantev:2005zs,Pantev:2005wj,Pantev:2005rh,Nussinov:2006iva,Nussinov:2008aa,Nussinov:2009zz, Nussinov:2011mz,Banks:2010zn,Distler:2010zg,Kapustin:2014gua}),
 where charged objects are extended objects rather than pointlike operators for ordinary symmetries. 
In particular, the Maxwell theory in $(3+1)$ dimensions can be understood as a broken phase of $U(1)$ 1-form symmetry whose charged object is a Wilson loop, and the photon can be identified 
as the NG mode~\cite{Kugo:1985jc,Kovner:1992pu,Gaiotto:2014kfa,Lake:2018dqm}. The counting rule of the number of NG modes for higher-form symmetries has also been formulated recently~\cite{Hidaka:2020ucc}.

On the other hand, there are cases where NG modes become unstable in the presence of background fields, which are beyond the conventional counting rule above. In fact, such instabilities have been discussed in various contexts, such as the instability of a photon with a time-dependent axion background in cosmology~\cite{Carroll:1989vb,Garretson:1992vt,Anber:2006xt}
or with a chirality imbalance known as the chiral plasma instability in cosmology and astrophysics~\cite{Joyce:1997uy,Akamatsu:2013pjd}, and instability of an emergent dynamical axion with the background electric field in condensed matter physics~\cite{Ooguri:2011aa}; see also  Ref.~\cite{Nakamura:2009tf} for a related instability in the five-dimensional Maxwell-Chern-Simons theory with a constant electric field.
One can ask whether the existence of such instabilities may be understood as a universal property of NG modes dictated by some symmetry algebra and whether there may be a general counting rule for these unstable modes similar to that of usual NG modes.

In this paper, we derive a general counting rule of these unstable NG  modes for the spontaneous breaking of internal symmetries in the presence of background fields.
We show that the number of unstable NG modes is determined by the rank of the matrix in terms of the correlation functions of broken  symmetry generators; see \er{main} for our main result. 
We verify the validity of our formula for known examples of  instabilities.

\section{Comparison between type-B and unstable Nambu-Goldstone modes}
Before going to the detailed discussion, we briefly summarize the previous studies and our results on the classification of NG modes based on the dispersion relations.
In Lorentz-invariant systems, the dispersion relation of NG modes is always $\omega = |\bs{k}|$, and the number of NG modes is equal to the number of broken symmetry generators~\cite{Nambu:1961tp,Goldstone:1961eq,Goldstone:1962es}. 
In the presence of background fields where Lorentz symmetry is explicitly broken, it has been shown that there can be NG modes with the quadratic dispersion $\omega \sim |\bs{k}|^2$ and gapped modes with $\omega = {\rm const} + O (|\bs{k}|^2)$ for 0-form symmetries~\cite{Nielsen:1975hm} and higher-form symmetries~\cite{Yamamoto:2015maz,Sogabe:2019gif}. 
The counting rules of these modes for 0-form symmetries and higher-form symmetries are proved in Refs.~\cite{Watanabe:2012hr,Hidaka:2012ym} and Ref.~\cite{Hidaka:2020ucc}, respectively. 
Without fine-tuning of parameters of a theory, whether the dispersion relation of a NG mode is linear or quadratic dispersion is classified by a quantity $\rho_{\cal IJ}^0 \equiv \langle [Q_{\cal I}, Q_{\cal J}] \rangle$ in the ground state. Here, $Q_{\cal I}$ are broken symmetry generators whose independent degrees of freedom are parametrized by the index ${\cal I}$ (see Ref.~\cite{Hidaka:2020ucc} for the detailed definition), and the superscript ``0'' refers to the temporal direction in which commutators are defined: NG modes for $\rho_{\cal IJ}^0 = 0$ have linear dispersion and are called type A, while NG modes for $\rho_{\cal IJ}^0 \neq 0$ have quadratic (or gapped) dispersion and are called type B.

Our new insight in this paper is that there are generically additional modes with the dispersion $\omega \sim \pm {\rm i} |\bs{k}|$ that can also be understood as NG modes dictated by the symmetry algebra. Although several examples of the modes with this dispersion relation with the positive imaginary part are already known simply as instabilities~\cite{Carroll:1989vb,Garretson:1992vt,Joyce:1997uy,Ooguri:2011aa,Akamatsu:2013pjd,Anber:2006xt}, they have not been identified as NG modes so far.%
\footnote{The partner mode with the negative imaginary part, $\omega \sim - {\rm i} |\bs{k}|$, also has a remarkable feature: Although it is a damping mode, it does not involve any entropy production (somewhat similarly to Landau damping) unlike usual diffusion modes with the dispersion $\omega \sim - {\rm i} |\bs{k}|^2$. While it can also be understood as a new type of NG mode, below we will mostly focus on the unstable NG mode with the positive imaginary part, as the number of the former is simply equal to the number of the latter. This is similar to  the fact that type-B NG modes are accompanied by gapped modes, and their numbers are equal.}
Moreover, such a mode for even the conventional 0-form symmetry has not been known to the best of our knowledge. 
The purpose of this paper is to generalize the counting rule for the conventional NG modes to these unstable NG modes and to provide a new such example of 0-form symmetry. To this end, we will introduce a new quantity $\rho^l_{\cal IJ}$ in Eq.~(\ref{rho_cal}) below, which is the matrix of the correlators of broken symmetry generators put on the planes perpendicular to the spatial $x^l$ direction. The comparison between type-B and unstable NG modes is summarized in Table~\ref{tab:classification}. 

\begin{table}
    \begin{tabular}{c|c|c}
      & Dispersion & Condition  \\
      \hline 
      Type B & $\omega \sim  |\bs{k}|^2$ & $\rank \rho^0_{\cal IJ} \neq 0$  \\
      \hline
       Unstable & $\omega \sim {\rm i} |\bs{k}| $ & 
       $\rank \rho^l_{\cal IJ} \neq 0$ 
    \end{tabular}
    \caption{Comparison between type-B and unstable NG modes. Here, $\rho^0_{\cal IJ}$ is the matrix of the equal time commutators of the broken symmetry generators put on spatial directions, and $\rho^l_{\cal IJ}$ is the matrix of the correlators of broken symmetry generators put on the planes perpendicular to the spatial $x^l$ direction.}
    \label{tab:classification}
\end{table}

Before going to a mathematical proof for a more generic case, we first provide a rough idea on when and how the type-B or unstable NG modes appear. In essence, the directions of background fields can be classified to spatial and temporal ones (whose precise definitions will be given in Sec.~\ref{sec:counting} below). All previous works on the classification of NG modes implicitly assume that background fields are in the spatial direction.
In this case, the dispersion relation is modified by the linear term of $\omega$ as $\omega^2  = \pm \alpha \omega + |\bs{k}|^2 +\cdots $ with $\alpha$ being some real constant, leading to type-B NG modes and gapped modes. The number of type-B NG modes can be counted by the correlation of spatially extended symmetry generators.

On the other hand, if the directions of the background field strengths are temporal, the modification of the dispersion relation is given by the linear term of $|\bs{k}|$ as $\omega^2  = \pm \beta |\bs{k}| + |\bs{k}|^2$ with $\beta$ being some real constant, leading to unstable NG modes and dumping modes. 
In this case, the number of unstable NG modes can be counted by the correlation of temporally extended symmetry generators.

In the following, we put this argument on a more mathematical basis using effective field theories for higher-form symmetries.

\section{Effective theories}%
We consider low-energy effective theories for spontaneous breaking of continuous 0- and higher-form internal symmetries with couplings to 
background fields.
To discuss the dispersion relations in the low-energy region, it is sufficient to focus on the effective action up to second order in derivatives. 
Note that, up to second order in fields and derivatives, the effective action for the higher-form symmetries includes that of 0-form symmetries.
We consider $D$-dimensional Minkowski spacetime with the mostly plus metric $\eta_{\mu\nu} = \diag (-1,1,...,1)$.

For the 0-form symmetries, we assume that a continuous 0-form symmetry with a compact Lie group $G$ is spontaneously broken to its subgroup $H$. For the higher-form symmetries, we assume that $U(1)$ $p_I$-form symmetries ($I = 1,..., N$) are spontaneously broken, since higher-form symmetries are always Abelian~\cite{Gaiotto:2014kfa}.
We introduce a charged object for the $p_I$-form symmetry, which is a Wilson loop $W(C_I) = \exp \left(\rmi \int_{C_I }a_I \right)$ on a $p_I$-dimensional closed subspace $C_I$. Here, $a_I$ is a $p_I$-form field, which has a gauge redundancy under the transformation $ a_I \to a_I + \rmd \lambda_I $ for a $U(1)$ $(p_I-1)$-form parameter $\lambda_I$ satisfying $ \int_{C_I} \rmd \lambda_I \in 2\pi \bb{Z}$. 
For a $U(1)$ symmetry, the Wilson loop is transformed as $W_I (C_I ) \to \rme^{\rmi \alpha_I} W(C_I)$ with $\rme^{\rmi \alpha_I} \in U(1)$, which is generated by a shift $a_I \to a_I + \epsilon_I$ with a $p_I$-form $\epsilon_I$ satisfying $\rmd \epsilon_I =0$ on $C_I$ and  $\int_{C_I} \epsilon_I = \alpha_I$.
For a 0-form symmetry with a non-Abelian group, the global transformation leads to a constant shift for the leading order of a field.
The spontaneous breaking of the higher-form symmetries can be characterized by the nonzero vacuum expectation value of the Wilson loop in the large volume limit, $\vevs{W(C_I)} \to 1$ up to the renormalization that can depend on the volume of $C_I$.

We now construct the effective action. To have the action preserving higher-form global symmetries, we use the field strength  $f_{I} = \rmd a_I$. 
Up to the second order of the derivatives and fields, the effective action is given by%
\footnote{For a construction of the effective actions based on the nonlinear realization of higher-form symmetries, see Ref.~\cite{Hidaka:2020ucc}.}
\begin{equation}
S 
 =
- \fr{1}{2} \int F^2_{IJ} \rmd a_I \wed *\rmd a_J
+
\fr{1}{2}
\int \rmd a_I \wed \rmd a_J \wed  A_{IJ} \,.
\label{S}
\end{equation}
Here, $F^2_{IJ} = F_{KI} F_{KJ}$ is expressed in terms of an invertible matrix $F_{IJ}$ that represents the matrix of decay constants, and 
$A_{IJ} = (-1)^{(p_I+1)(p_J+1) }A_{JI}$ is a $p_{IJ}$-form with $p_{IJ} = D- p_I -p_J -2$ that can depend on the spacetime coordinate.%
\footnote{The background field satisfies the quantization condition $\int_{\Sigma_{IJ}} \rmd A_{IJ} \in \fr{1}{2\pi} \bb{Z}$ on a $(p_{IJ}+1)$-dimensional closed compact subspace $\Sigma_{IJ}$.} 
The first term in \er{S} can be understood as a generalization of the Maxwell term, and the second term is a topological term with a possible background field $A_{IJ}$.
The matrix $F_{IJ}$ is nonzero only for $p_I = p_J$.
It would also be possible to regard $F_{IJ}^2$ as a $(p_J - p_I)$-form background field, but we assume that $F_{IJ}$ is a just constant 0-form for simplicity.
The presence of the background field $A_{IJ}$ breaks Lorentz invariance in general.
Note that the assumption of the $p_I$-form global symmetries of the action excludes terms that break these symmetries explicitly, such as dynamical charged matter, which may be allowed to exist if we assume only $(p_I -1)$-form gauge invariance.

In passing, we remark that we can reproduce the nonrelativistic effective actions previously considered in Refs.~\cite{Watanabe:2012hr, Hidaka:2012ym, Watanabe:2014fva} by choosing, e.g., $A_{IJ} = \mu_{IJ} x^1 \rmd x^2\weds \rmd x^{D-1} $ with an appropriate constant $\mu_{IJ}$ up to a total derivative.
It is possible to identify $A_{IJ}$ as a background gauge field whose symmetry is called composite or Chern-Weil global symmetry~\cite{Brauner:2020rtz,Heidenreich:2020pkc}.

\section{Counting rule of unstable NG modes}%
\label{sec:counting}
To count the number of the unstable modes in the presence of the background field, we first derive the dispersion relations.
Hereafter, we assume that the translational invariance of the system is not broken.
We consider the configuration of the background field in the temporal direction where $\rmd A_{IJ} = \fr{1}{p_{IJ}!} E_{IJ,i_1...i_{p_{IJ}}} \rmd x^0 \wed \rmd x^{i_1} \weds \rmd x^{i_{p_{IJ}}}$ with a constant $E_{IJ,i_1...i_{p_{IJ}}}$.
The background field can be understood as a generalization of an ordinary background electric field.%
\footnote{If we introduce a background magnetic field or its generalization in the spatial direction instead of the electric field, some of the NG modes become gapped rather than unstable~\cite{Yamamoto:2015maz,Brauner:2017mui,Sogabe:2019gif,Hidaka:2020ucc}.}

\subsection{Dispersion relations}
The equation of motion for $a_I$ is
\begin{equation}
F^2_{IJ} \rmd * f_J
= (-1)^{p_J +1} 
f_J  \wed \rmd A_{IJ}.
\label{EOM}
\end{equation}
To focus on physical degrees of freedom, we take the temporal gauge $a_{I,0 i_1...i_{p_I-1}} =0$ with the Gauss law constraint $\der^i f_{I,0 i i_1 ... i_{p_I-2}} =0$.
In momentum space,
\er{EOM} 
can be written as
\begin{equation}
(\omega^2 - |\bs{k}|^2)
 \hat{a}_I^{i_1...i_{p_I}}
 = -\fr{\rmi}{p_J!}
\hat{M}^{ i_1...i_{p_I} l  j_1 ... j_{p_J}}_{IJ}
k_l \hat{a}_{J, j_1 ...j_{p_J}}\,,
\label{EOM_momentum}
\end{equation}
where we defined
$\hat{a}_I = F_{IJ} a_J $
to simplify our notation.
We also introduced $\hat{M}^{i_1...i_{p_I} l j_1 ... j_{p_J}}_{IJ}
 = 
 F^{-1}_{KI} M^{i_1...i_{p_I} l j_1 ... j_{p_J}}_{KL} F_{LJ}^{-1}$, where
\begin{equation}
M^{i_1...i_{p_I} l j_1 ... j_{p_J}}_{IJ}
= 
\fr{\epsilon^{0  i_1... i_{p_I} l j_1 ..j_{p_J}
k_1 ... k_{p_{IJ}}}}{p_{IJ}!
} 
E_{IJ, k_1 ... k_{p_{IJ}}}\,, 
\end{equation}
with $\epsilon_{\mu_1...\mu_D}$ being the totally antisymmetric tensor satisfying $\epsilon_{01...D-1} = 1$.
We remark that $\hat{M}^{ i_1...i_{p_I}l  j_1 ... j_{p_J}}_{IJ}$ 
is antisymmetric under the exchange 
${\cal I} = (I, i_1,...,i_{p_I}) \corr {\cal J} = (J, j_1,..., j_{p_J})$.
Using these collective indices, the equation of motion in \er{EOM_momentum} can be simplified as
\begin{equation}
(\omega^2 - |\bs{k}|^2)
 \hat{a}_{\cal I}
 = 
 - \rmi k_l \hat{M}_{{\cal IJ}}^l
 \hat{a}_{\cal J},
\label{EOM_collective}
\end{equation}
with the antisymmetric matrices 
$\hat{M}_{{\cal IJ}}^l = \hat{M}^{i_1...i_{p_I} l j_1 ... j_{p_J}}_{IJ}$
satisfying 
$\hat{M}_{\cal JI}^l  = -   \hat{M}_{\cal IJ}^l$.
Here, the indices $(i_1,...,i_{p_I})$ in ${\cal I}$ are ordered as $i_1< \cdots < i_{p_I}$ to avoid overcounting.
The number of degrees of freedom of ${\cal I}$ is denoted as ${\cal N}$.

We can now count the number of unstable modes in the background 
$\hat{M}_{{\cal IJ}}^l \neq 0$. Equation~\eqref{EOM_collective} implies that the dispersion relations may depend on the direction of the wave vector. 
We thus consider the dispersion relation for each direction.
If we choose a wave vector along the $x^l$ direction, i.e., $\bs{k} = (0,...0,\underbrace{k}_{l\text{-th}},0,...,0)$, \er{EOM_collective} is further simplified as
 $(\omega^2 - k^2)
 \hat{a}_{\cal I}
 = 
- \rmi k \hat{M}_{{\cal IJ}}^l
 \hat{a}_{\cal J}$.
Since the matrix $\hat{M}^l = (\hat{M}_{\cal IJ}^l)$ is antisymmetric, we can transform this matrix by using an orthogonal matrix $P$ into the form
\begin{equation}
\begin{split}
&
P \hat{M}^{l} P^T 
= \mtx{ \Lambda_{1}^l && \\ &\ddots& \\ && \Lambda_{n }^l \\ &&&
0_{({\cal N} - 2n)\times ({\cal N} -2n)}
}\,,
\\
& 
 \Lambda_{ m}^l  = \mtx{ 0 & - \lambda_{m}^l  \\ \lambda_m^{l} & 0}\,, 
\quad 
n = \fr{1}{2}\rank (\hat{M}^{l})\,.
\end{split}
\end{equation}
Note that $n$ is an integer because the rank of an antisymmetric matrix is always an even integer.
In this basis, we can solve \er{EOM_collective} and obtain the dispersion relation for each $\Lambda_{m }^l$:
\begin{equation}
\label{instability}
\omega^2 = k^2 \pm |k\lambda_{m}^l|,
\end{equation}
which exhibits an instability in the region $|k| < |\lambda_{m}^l|$.
The number of the unstable NG modes $N^l_{\rm unst}$ in the $x^l$ direction is, therefore,
\begin{equation}
    N^l_{\rm unst} 
    =
    \fr{1}{2}\rank (\hat{M}_{{\cal IJ} }^l)=\fr{1}{2}\rank (M_{{\cal IJ} }^l)\,.
    \label{Nunst}
\end{equation}
We remark that the Gauss law $k_i a^{i i_1 ...i_{p_I -1}} =0 $ is automatically satisfied for the counting of the unstable modes, since $ k_i M_{{\cal IJ} }^i = 0$ if the direction of $k_i$ is the same as either of polarization directions, $i_1,...,i_{p_I}$ or $j_1,..., j_{p_J}$, due to the totally antisymmetric tensor.
In other words, the unstable NG modes are always transverse modes.
Note that the number of unstable modes is determined for a given direction of the wave vector.

\subsection{Correlation function between symmetry generators}%
Here, we show that the matrix $M_{{\cal IJ} }^l$ can be expressed by the correlation function of broken symmetry generators, and, hence, the number of the unstable NG modes is equal to half of the rank of the matrix in terms of the correlation function (see the Appendix for detail).

From the equation of motion in \er{EOM}, we can define a conserved charge on a $(D - p_I - 1)$-dimensional closed subspace $\Sigma_{I}$:
\begin{equation}
 Q_I (\Sigma_{I})
 = \int_{\Sigma_{I}} 
\left(- F^{2}_{IJ} *  f_J + 
f_J \wed {A}_{IJ}\right) 
.
\end{equation}
The correlation function can be calculated in the path integral formulation by using Ward-Takahashi identity as
\begin{equation}
\rmi \vevs{ Q_I (\Sigma_{I})  Q_J (\Sigma_{J})}
= - 
\int_{\Sigma_I \cap \Omega_J}
 \rmd A_{IJ}\,,
\label{correlation}
\end{equation}
where $\Omega_J$ is a $(D-p_J)$-dimensional subspace satisfying $\der \Omega_J = \Sigma_J$. 
From the correlation function, we can extract the matrix $M^l_{\cal IJ}$
by using a spatial version of the equal-time commutation relation of symmetry generators, where the $x^l$ direction plays the role of the temporal direction.
A $(D- p)$-dimensional plane localized at $x^{i_1} = \cdots = x^{i_p} =0$ is denoted by $S^{i_1...i_p}$.
To have a commutation relation, we also introduce a plane $S^{i_1...i_p}_{x^{i_p} = c}$ localized at $x^{i_1} = \cdots = x^{i_{p-1}} =0$, $x^{i_p} = c$ with $c$ being some constant.
We first take the large volume limit $\Sigma_I \to S^{i_1...i_{p_I} l} \cup \b{S}^{i_1...i_{p_I} l}_{x^l = - \epsilon}$ for an infinitesimal positive parameter $\epsilon$. Here, $S^{i_1...i_{p_I} l}$ intersects with $\Omega_J$ while $\b{S}^{i_1...i_{p_I} l}_{x^l = - \epsilon}$ (which is $S^{i_1...i_{p_I} l}_{x^l = - \epsilon}$ with an opposite orientation) does not.
Then, we take the large volume limit 
$\Sigma_J \to 
S^{j_1...j_{p_J} l}_{x^l =\fr{\epsilon}{2} } \cup 
\b{S}^{j_1...j_{p_J} l}_{x^l = - \fr{\epsilon}{2} }$.
In these limits, we have 
$\Sigma_I \cap \Omega_J \to
S^{i_1...i_{p_I} l} \cap 
S^{j_1...j_{p_J} l} = 
S^{i_1...i_{p_I} l j_1...j_{p_J}}$,
and, hence,
\begin{align}
&
\fr{\rmi \vevs{ Q_I (\Sigma_{I})  Q_J(\Sigma_{J})}}{\vol (\Sigma_I \cap 
\Omega_J)} 
\nonumber \\
&\to
\fr{\rmi \vevs{ Q_I (S^{i_1...i_{p_I} l} \cup 
\b{S}^{i_1...i_{p_I} l}_{x^l = - \epsilon} )  Q_J(S^{j_1...j_{p_J} l}_{x^l =\fr{\epsilon}{2} } \cup 
\b{S}^{j_1...j_{p_J} l}_{x^l = - \fr{\epsilon}{2} })}}{\vol (S^{i_1...i_{p_I} l j_1...j_{p_J}})} 
\nonumber \\
&
=:
\rho^{i_1...i_{p_I} l j_1...j_{p_J}}_{IJ}
\,.
\label{rho}
\end{align}
Here, $\vol (\Sigma)$ is the volume of a subspace $\Sigma$ including the temporal direction.
By the explicit calculation of \er{correlation}, we have
\begin{align}
&
\rho^{i_1...i_{p_I} l j_1...j_{p_J}}_{IJ}
=(-1)^{p_{IJ} (D-p_{IJ}-1)+1}
M^{i_1...i_{p_I} l j_1...j_{p_J}}_{IJ}\,.
\label{rhoM}
\end{align}
Combining \ers{Nunst} and \eqref{rhoM}, we arrive at 
\begin{equation}
    N^l_{\rm unst}=\frac{1}{2}\rank (\rho^l_{\cal IJ}) \,,
    \label{main}
\end{equation}
where we have defined 
\begin{equation}
\label{rho_cal}
\rho^l_{\cal IJ}  
=
\rho^{i_1...i_{p_I} l j_1...j_{p_J}}_{IJ}.
\end{equation}

Note that our counting rule in \er{main} is similar to that of the type-B NG modes for the higher-form symmetries derived in Ref.~\cite{Hidaka:2020ucc}.
The difference between the counting of type-B NG modes and unstable modes is whether the directions of the nonzero components of field strengths for the background fields are spatial directions $\rmd A_{IJ} \sim \rmd x^{i_1} \weds \rmd x^{i_{p_{IJ}+1}} $ or temporal directions $\rmd A_{IJ} = \fr{1}{p_{IJ}!}E_{IJ, i_1...i_{p_{IJ}}} \rmd x^0 \wed \rmd x^{i_1} \weds \rmd x^{i_{p_{IJ}}}$.

We also note that the correlation function evaluated here indicates only the presence of instabilities rather than the presence of long-lasting ones.
Although we assumed $\rmd A_{IJ}$ to be a constant when evaluating the correlation function, the background field may evolve dynamically to relax the instabilities in realistic systems. It is indeed the case for the chiral plasma instability discussed in Refs.~\cite{Joyce:1997uy, Akamatsu:2013pjd}.

\section{Examples}
In this section, we provide examples of unstable NG modes to verify the validity of our general counting rule in physical systems.

\subsection{Chiral plasma instability}%
First, we consider electromagnetism in the background of an axion field:
\begin{equation}
 S
 =
-\fr{1}{2e^2} \int \rmd a \wed *\rmd a  
+ 
 \fr{1}{2}\int \Theta_{aa} \rmd a \wed \rmd a\,. 
\end{equation}
Here, the photon is described by a 1-form gauge field $a = a_\mu \rmd x^\mu$, which can be understood as a NG mode for the spontaneously broken $U(1)$ 1-form symmetry, $e$ is a coupling constant, and $\Theta_{aa} = -2 C \mu_5 x^0$ is the background axion field with $\mu_5$ corresponding to the chiral chemical potential for $C = 1/(4\pi^2)$. In fact, the second term above can be rewritten as the effective Chern-Simons term $S_{\rm CS} = -C \int \mu_5 \epsilon^{0ijk} a_i \der_j a_k$~\cite{Redlich:1984md} by integration by parts. Then, taking its variation with respect to $a_i$ reproduces the so-called chiral magnetic effect: $j^i := e^2\delta S_{\rm CS}/\delta a_i = -2C e^2\mu_5 \epsilon^{0ijk} \der_j a_k$~\cite{Vilenkin:1980fu,Nielsen:1983rb,Fukushima:2008xe}.

The equation of motion for the photon is $\fr{1}{e^2} 
\square 
a^i 
- 
2C \epsilon^{0 i j k} 
\mu_5  \der_j a_k=0,
$
where we have taken the temporal gauge $a_0 =0$ and the Gauss law constraint $\der_i a^i =0$. Under the plane wave ansatz, we have the equation of motion in momentum space:
$
(\omega^2 - |\bs{k}|^2) \hat{a}^i 
+
 \rmi
 \hat{M}^{ilj} k_l
 \hat{a}_j=0,
$
where we defined $\hat{a}_\mu = \fr{1}{e}a_\mu$ and 
$\hat{M}^{ilj}
=e^2 M^{ilj}  = 
-
2 \epsilon^{0ilj} C e^2 \mu_5 $.
For instance, we choose the wave vector along the $x^1$ direction, $\bs{k} = (k_1, 0,0)$.
In this case, the Gauss law constraint leads to $a_1=0$ for nonzero $k_1$.
Then the equation of motion is simplified as
\begin{equation}
 \rmi  k_1 
\mtx{
0 &
 \hat{M}^{213 } 
\\
 \hat{M}^{312}
& 0
}\mtx{\hat{a}_2 \\ \hat{a}_3}
= 
(\omega^2 - k_1^2)
\mtx{\hat{a}_2 \\ \hat{a}_3}.
\end{equation}
The dispersion relation is obtained as
$\omega^2 
= k_1^2  \pm 
| \hat{M}^{213 }
 k_1|
$.
Therefore, we have one unstable mode in the $x^1$ direction (and similarly for the $x^{2,3}$ directions). 
This is the chiral plasma instability~\cite{Joyce:1997uy,Akamatsu:2013pjd}.

Next, we discuss the relation between the matrix $M^{ilj} $ and the correlation function of symmetry generators.
The equation of motion for $a$ gives the conserved charge on a closed surface $\Sigma_a$:
\begin{equation}
 Q (\Sigma_a)
 = 
\int_{\Sigma_a}\left(
-\fr{1}{e^2} * \rmd a
+
\Theta_{aa} \rmd a \right)\,.
\end{equation}
The correlation function between two symmetry generators is 
\begin{equation}
 \rmi \vevs{ Q (\Sigma_a)  Q (\Sigma'_a)}
 =
-
\int_{\Sigma_a \cap \Omega_a' }\rmd \Theta_{aa}\,,
\end{equation}
where $\Omega_a'$ is a world volume whose boundary is a closed surface $\Sigma_a'$.
Now, we take the limit 
$\Sigma_a \to S^{il} \cup \b{S}^{il}_{x^l = -\epsilon} $
and then 
$\Sigma_a' \to S^{jl}_{x^l = \epsilon/2} \cup \b{S}^{jl}_{x^l = -\epsilon/2} $.
We have a temporally extended one-dimensional subspace 
$\Sigma_a \cap \Omega_a' \to  S^{ilj}$. 
Then, the matrix $\rho^{ilj}$ in \er{rho} becomes
\begin{equation}
\begin{split}
& 
\rho^{ilj}
 = 
\fr{2C \epsilon^{0i l j} \int_{S^{ilj}} \mu_5 \rmd x^0 }{\vol(S^{ilj})}
= - M^{ilj}\,. 
\end{split}
\end{equation}
The number of the unstable NG modes along the $x^l$ direction coincides with $\fr{1}{2}\rank (\rho^{ilj})$.

\subsection{Dynamical axion in electric field}%
The next example is the dynamical axion in the background electric field in $(3+1)$ dimensions~\cite{Ooguri:2011aa}. We consider the effective action
\begin{equation}
 S = -\fr{F^2_\phi}{2} \int |\rmd \phi|^2 -\fr{1}{2e^2}
\int |\rmd a|^2
+ 
\int \rmd\phi \wed \rmd a \wed A_{\phi a}\,,
\end{equation}
where $\phi$ is a 0-form axion field, $F_\phi$ is a decay constant of  the axion, and $\rmd A_{\phi a} = E_{\phi a, i} \rmd x^0 \wed \rmd x^i$ with a constant $E_{\phi a,i }$.
In the plane wave basis, the equations of motion for $\phi$ and $a$ can be written as
$  (\omega^2 -|\bs{k}|^2) \hat{\phi }
+
\rmi k_l \hat{M}^{lj}_{\phi a}
  \hat{a}_j 
=0$
and 
$(\omega^2 -|\bs{k}|^2) \hat{a}^i 
+
\rmi k_l 
\hat{M}^{i l}_{ a \phi}
\hat\phi 
=0$,
respectively.
Here, we defined
$\hat\phi = F_\phi \phi$,
 $\hat{a}_\mu = \fr{1}{e} a_\mu$, and 
 $\hat{M}^{li}_{\phi a} = - \hat{M}^{il}_{ a \phi }
  = \fr{e} {F_{\phi}}M^{li}_{\phi a} = 
  \epsilon^{0 lik}\fr{e} {F_{\phi}} E_{\phi a,k }$.
For concreteness, we take our coordinate so that 
$E_{\phi a,i } = (0,0,E_{\phi a,3 })$,
and we focus on the wave vector $\bs{k} = (k_1, 0,0)$.
We then find the dispersion relations $ \omega^2 = k_1^2$, $ k_1^2 \pm |k_1 \hat{M}_{\phi a}^{12}|$, among which there is one unstable mode.

We can also find the nontrivial correlations between symmetry generators. The equations of motion for $\phi$ and $a$ lead to the conserved charges
\begin{equation}
\begin{split}
 Q_{\phi} (\Sigma_\phi )
&
=  \int_{\Sigma_\phi} \left(-F^2_\phi *\rmd \phi +
\rmd a  
\wed A_{\phi a}\right),
\\
 Q_{a} (\Sigma_a)
&
= \int_{\Sigma_a} \left(-\fr{1}{e^2}* \rmd a + 
\rmd\phi \wed A_{\phi a}\right).
\end{split}
\end{equation}
Here, $\Sigma_\phi$ is a three-dimensional closed subspace.
The correlation function between them is
\begin{equation}
\rmi \vevs{  Q_{\phi} (\Sigma_\phi)
Q_{a} (\Sigma_a)}
 =  - 
\int_{\Sigma_\phi  \cap \Omega_a} \rmd A_{\phi a}\,.
\end{equation}
We take the limit 
$\Sigma_\phi \to S^{l} \cup \bar{S}^{l}_{x^l = - \epsilon}$ 
and then the limit 
$\Sigma_a \to S^{jl}_{x^l =\epsilon/2} \cup \b{S}^{jl}_{x^l = -\epsilon/2}$.
In this case, we have 
$\Sigma_\phi  \cap \Omega_a
\to  S^{lj}$, and
\begin{equation}
\begin{split}
&
\rho_{\phi a}^{lj}
= 
- M^{lj}_{\phi a}
\,.
\end{split}
\end{equation}
For the above choice of $(E_{\phi a , i})$, we have, e.g.,
$\fr{1}{2}\rank (\rho^1) = 1$, 
which matches the number of unstable NG modes propagating along the $x^1$ direction.

\subsection{Unstable NG modes from 0-form symmetry breaking}
Based on our generic description of unstable NG modes above, we give the third example of an unstable NG mode for a conventional 0-form symmetry with a background vector field, which is simple yet new to the best of our knowledge.

We consider the following low-energy effective action for the spontaneous breaking of the 0-form symmetry $U(1)\times U(1) \to \{1\}$ in $(2+1)$ dimensions:
\begin{equation}
 S = -\fr{F_\phi^2}{2} 
\int |\rmd\phi|^2 - \fr{F_\chi^2}{2} \int |\rmd\chi|^2
+ 
\int \rmd\phi \wed \rmd\chi \wed A_{\phi \chi}\,.
\label{S_0}
\end{equation}
Here, $(\rme^{\rmi \phi }, \rme^{\rmi\chi})$ is the set of 
NG modes for the $U(1)\times U(1)$ symmetry,
and 
$A_{\phi \chi} = - A_{\chi \phi}
=  A_{\phi \chi, \mu} 
\rmd x^\mu 
= E_{\phi \chi, i } x^0 \rmd x^i $
is a background field with a constant $E_{\phi \chi, i}$.
One may understand $A_{\phi \chi}$ as a background gauge field whose conserved charge is the number of linked vortex loops for $\phi $ and $\chi$~\cite{Brauner:2020rtz}.
The equations of motion for $\phi$ and $\chi $ in the plane wave basis  are
$  (\omega^2 - |\bs{k}|^2) \hat{\phi}
+
\rmi k_l \hat{M}^l_{\phi \chi} \hat\chi 
=0$
and
$ (\omega^2 - |\bs{k}|^2) 
\hat\chi
-
\rmi k_l \hat{M}^l_{\phi \chi} \hat\phi  =0$,
respectively.
Here, we defined 
$\hat{\phi} = F_\phi \phi$, 
$\hat{\chi} = F_\chi \chi$,
and
$ 
\hat{M}_{\phi \chi}^l
= 
\fr{1}{F_\phi F_\chi} M_{\phi \chi}^l 
= \fr{1}{F_\phi F_\chi}
\epsilon^{0  l i } E_{\phi \chi, i }$. 
Without loss of generality, we choose the direction of the background field so that only $E_{\phi \chi , 2}$ is nonvanishing.
In this case, we have 
$M^1_{\phi \chi} 
=
\epsilon^{0 12} E_{\phi \chi, 2}
=
- E_{\phi \chi, 2}
$
and $M^2_{\phi \chi} =0$.
We also choose a wave vector 
$\bs{k} = (k_1,0)$.
Then, we have the dispersion relation
$\omega^2 =  k_1^2
\pm 
|k_1 \hat{M}_{\phi \chi}^1|$, 
and there exists one unstable mode for
$|k_1| < |\hat{M}_{\phi \chi}^1|$ along the $x^1$ direction.

Next, we relate the matrix $M_{\phi \chi}^l $ to the correlation function of symmetry generators. The conserved charges for the spontaneously broken 0-form symmetries are found from the equations of motion as
\begin{equation}
\begin{split}
 Q_\phi (\Sigma_\phi)
 &= \int_{\Sigma_\phi} 
 (-F_\phi^2 *\rmd\phi 
+
\rmd\chi \wed A_{\phi \chi})\,,
\\
Q_\chi (\Sigma_\chi)
 &
 =
 \int_{\Sigma_\chi} 
 (-F_\chi^2 *\rmd\chi 
-
\rmd\phi \wed A_{\phi \chi} )\,,
\end{split}
\end{equation} 
where $\Sigma_{\phi, \chi}$ are closed surfaces. 
The conserved charges satisfy the correlation function
\begin{equation}
 \rmi \vevs{Q_\phi(\Sigma_{\phi}) Q_\chi(\Sigma_{\chi})}
 = 
-\int_{\Sigma_\phi \cap \Omega_\chi}  \rmd A_{\phi \chi} \,.
\end{equation}
Here,
$\Omega_\chi$ is
a three-dimensional subspace whose boundary is $\Sigma_\chi$.
To have the matrix $M_{\phi \chi}^l$, we take the limit 
$\Sigma_\phi \to S^l\cup \bar{S}^l_{x^l  = -\epsilon}$
and then the limit
$\Sigma_\chi \to 
S^l_{x^l  = \epsilon/2}\cup \bar{S}^l_{x^l  = -\epsilon/2}$.
We have 
$\Sigma_\phi \cap \Omega_\chi \to S^l $ and
\begin{equation}
\begin{split}
\rho_{\phi \chi}^l
= M^l_{\phi \chi} \,.
\end{split}
\end{equation}
For the $x^1$ direction, we have $\fr{1}{2} \rank (\rho^1) = 1$, which 
coincides with the number of unstable NG modes.

\sectaps{Discussions}%
We remark that our derivation is similar to that of the counting rule for the so-called type-B NG modes~\cite{Watanabe:2012hr,Hidaka:2012ym,Watanabe:2014fva,Hidaka:2020ucc}. 
In our derivation, the background fields with only spatial directions 
$\rmd A_{IJ} \sim \rmd x^{i_1} \weds \rmd x^{i_{p_{IJ}+1}} $ of the latter is replaced by those with the temporal direction $\rmd A_{IJ} = \fr{1}{p_{IJ}!}E_{IJ, i_1...i_{p_{IJ}}} \rmd x^0 \wed \rmd x^{i_1} \weds \rmd x^{i_{p_{IJ}}} $, and the commutation relation of symmetry generators, i.e., the time-ordered product of symmetry generators, is replaced by the product ordered in the spatial directions.
One main difference here is that there are various choices of spatial directions, which leads to the fact that the unstable NG modes depend on the direction of the wave vector.
The classification of conventional type-B NG modes and unstable NG modes in the presence of both spatial and temporal background fields is deferred to future work. 
One may also be able to extend our counting rule for unstable NG modes to finite temperature in a way similar to the one for conventional NG modes of 0-form symmetries in Ref.~\cite{Minami:2015uzo}. In such a case, dissipative effects could modify the dispersion relation of unstable NG modes; see, e.g., Refs.~\cite{Joyce:1997uy,Akamatsu:2013pjd} in the case of chiral plasma instability.
Finally, it would also be interesting to explore possible physical realizations of the theory considered in \er{S_0} with straightforward extensions to $D$ dimensions by replacing $A_{\phi \chi}$ with a $(D-2)$-form field.

\section*{Acknowledgements}
This work is supported in part by the Keio Institute of Pure and Applied Sciences (KiPAS) project at Keio University and JSPS KAKENHI Grant No.~JP19K03852 (N.~Y.) and by JSPS KAKENHI Grants No.~JP21J00480 and No.~JP21K13928 (R.~Y.).

\clearpage
\appendix 

\onecolumngrid

\section{Derivation of correlation function in \er{correlation}}
Here, we derive the correlation function in \er{correlation}.
We use the path integral formulation, where the time-ordered product 
is manifest.
In this formulation, the correlation function can be written as
\begin{equation}
 \vevs{Q_I ({\Sigma_I})Q_J(\Sigma_J)} 
 = {\cal N} 
\int {\cal D}[a]
Q_I ({\Sigma_I})Q_J(\Sigma_J) \rme^{\rmi S},
\end{equation}
where ${\cal N}$ is the normalization factor such that $\vevs{1} = 1$ and ${\cal D}[a] = {\cal D}a_{p_1} \cdots {\cal D} a_{p_N}$ is the integral measure of the NG modes.
The subspace ${\Sigma_I}$ is $(D-p_I-1)$-dimensional closed subspace on which the conserved charge for the $p_I$-form symmetry is defined.

To evaluate this correlation function, we consider the leading order of the expansion of $\vevs{Q_J (\Sigma_J)}$ in $\delta a_I$. As the path integral is invariant under an infinitesimal change of the dynamical variable $a_I \to a_I + \delta a_I$, under the assumption that there are no anomalies, we have the Schwinger-Dyson equation
\begin{equation}
  \rmi \vevs{(\delta_{a_I}S) Q_J(\Sigma_J)}
= 
- \vevs{\delta_{a_I} Q_J(\Sigma_J)}.
\label{Schwinger-Dyson}
\end{equation}
Since infinitesimal symmetry transformations for the NG modes are shift transformations, the variation of the action in \er{S} gives the conserved current $j_I$ as
\begin{equation}
\begin{split}
 \delta_{a_I}S
 &
= \int 
\left(- F^2_{IJ} \rmd \delta a_I \wed *\rmd a_J 
+
\rmd\delta a_I \wed \rmd a_J \wed A_{IJ}\right)
=: (-1)^{(p_I+1)(D-p_I-1)} \int j_I \wed \rmd \delta a_I \,,
\end{split}
\end{equation}
where the conserved charge $Q_I(\Sigma_I)$ can be defined by 
$Q_I(\Sigma_I)
 = \int_{\Sigma_I} j_I$.
Therefore, the relation (\ref{Schwinger-Dyson}) can also be viewed as the Ward-Takahashi identity.
On the other hand, the variation of $Q_J(\Sigma_J)$ leads to 
\begin{equation}
 \delta_{a_I} Q_J (\Sigma_J) 
=
\int_{\Sigma_J} \rmd\delta a_I \wed A_{JI}
=
(-1)^{(p_I+1)(D - p_I- 1)}
\int_{\Sigma_J}  A_{IJ} \wed \rmd\delta a_I\,.
\end{equation}
Here, we have neglected the term $F_{JI}^2 \int_{\Sigma_J} * \rmd \delta a_I $, as it can be absorbed by the renormalization of the conserved charge (see Ref.~\cite{Cherman:2021nox} for technical details).
By taking $\rmd \delta a_I = \epsilon \delta_{p_I +1} (\Sigma_I)$ with an infinitesimal parameter $\epsilon$, we have the correlation between the charges:
\begin{equation}
\begin{split}
&  \rmi \vevs{Q_I (\Sigma_I) Q_J(\Sigma_J)}
= 
-
\int_{\Sigma_J} A_{IJ} \wed \delta_{p_I +1} (\Sigma_I)\,.
\end{split}
\label{correlation_suppl}
\end{equation} 
Here, we define the delta function $p$-form $\delta_p (\Sigma_{D-p})$ on a $(D-p)$-dimensional subspace $\Sigma_{D-p}$ such that the integral of a $(D-p)$-form $\omega_{D-p}$ is given by (see, e.g., Refs.~\cite{Chen:2015gma,Hidaka:2020izy})
\begin{equation}
 \int_{\Sigma_{D-p}} \omega_{D-p} 
 = \int \omega_{D-p} \wed \delta_{p} (\Sigma_{D-p})\,,
\end{equation}
which can be explicitly written as
\begin{equation}
     \delta_{p} (\Sigma_{D-p}
    )
= 
\fr{\epsilon_{\mu_1... \mu_{D-p} \nu_1...\nu_{p}}}{(D-p)!p!}
\left(\int_{\Sigma_{D-p}
} 
\rmd z^{\mu_1}\weds \rmd z^{\mu_{D-p}}
\delta^D (x-z) 
\right) 
\rmd x^{\nu_1} \weds \rmd x^{\nu_p}\,.  
\end{equation}
By the Stokes theorem, we can rewrite the right-hand side of \er{correlation_suppl} as
\begin{equation}
\begin{split}
\int_{\Sigma_J} A_{IJ} \wed \delta_{p_I +1} (\Sigma_I)
&
= \int_{\der \Omega_J} A_{IJ} \wed \delta_{p_I +1} (\Sigma_I)
= 
\int_{ \Omega_J} \rmd A_{IJ} \wed \delta_{p_I +1} (\Sigma_I) 
\\
&
=  
\int \rmd A_{IJ} \wed \delta_{p_I +1} (\Sigma_I)
\wed \delta_{p_J}( \Omega_J)
=  
\int_{\Sigma_I \cap \Omega_J} 
\rmd A_{IJ}\,.
\end{split}
\label{correlation_RHS}
\end{equation}
Here $ \Omega_J$ is a $(D-p_J)$-dimensional subspace whose boundary is $\Sigma_J$.

Now, we assume that $A_{IJ}$ gives the background electric field and its generalization,
$\rmd A_{IJ} = \fr{1}{p_{IJ}!}E_{IJ, i_1...i_{p_{IJ}}} \rmd x^0 \wed \rmd x^{i_1}\weds \rmd x^{i_{p_{IJ}}}$.
We extract the matrix $M_{\cal IJ}^l $ from the correlation function by taking the large volume limit of the subspaces of the symmetry generators.
In this limit, we have a spatial version of the equal-time commutation relation between symmetry generators, where we choose the $x^l$ direction to replace the temporal direction.
First, we take the limit where $\Sigma_I$ is a pair of temporally extended parallel planes, where either of the planes intersects with $\Omega_J$:
\begin{equation}
\begin{split}
&\Sigma_I
    \to
 S^{i_1...i_{p_I} l } 
 \cup 
 \bar{S}^{i_1...i_{p_I}l}_{x^l = - \epsilon}.
\end{split}
\end{equation}
Here, we define a $(D-p)$-dimensional planes $S^{i_1...i_p}$ and $S^{i_1...i_{p-1} l}_{x^l = c}$ as
\begin{equation}
    S^{i_1...i_{p}} 
    :=
    \{ (x^\mu) \in \bb{R}^D| 
    x^{i_1} = \cdots =  x^{i_{p}} =0 \},
    \quad
        S^{i_1...i_{p-1} l }_{x^l = c} 
    :=
    \{ (x^\mu) \in \bb{R}^D| 
    x^{i_1} = \cdots =  x^{i_{p-1}} =0, 
    x^l = c\},
\end{equation}
with $\epsilon$ being an infinitesimal positive number.
The planes are defined so that the delta function form is simplified as
\begin{equation}
\begin{split}
    \delta_{p} (S^{i_1...i_{p}} 
    )
 &
= 
\fr{\epsilon_{\mu_1... \mu_{D-p} \nu_1...\nu_{p}}}{(D-p)!p!}
\rmd x^{\nu_1} \weds \rmd  x^{\nu_p}
\int_{S^{i_1...i_{p}} 
} 
\rmd z^{\mu_1}\weds \rmd z^{\mu_{D-p}}
\delta^D (x-z)    
 \\
 &
 = 
\delta(x^{i_1}) 
\cdots
\delta(x^{i_p}) 
\rmd x^{i_1} \weds \rmd  x^{i_p}.
\end{split}
\end{equation}
Accordingly, the exchange of indices $(i_1,...,i_{p})$ is assumed to flip the orientation.
Hereafter, we assume that $S^{i_1...i_{p_I} l}$ intersects with $\Omega_J$ but $\b{S}^{i_1...i_{p_I}l}_{x^l = -\epsilon}$ does not.

Next, we take the limit of $\Sigma_J$, which is another pair of temporally extended parallel planes.
The directions of the normal vectors of these pairs contain the $x^l$ direction,
\begin{equation}
    \Sigma_J 
\to  S^{j_1...j_{p_J} l}_{x^l = \fr{\epsilon}{2}}  \cup 
 \bar{S}^{j_1...j_{p_J} l}_{x^l = -  \fr{\epsilon}{2}}.
\end{equation}
Under this choice,
$\Omega_J$ can be written as
\begin{equation}
    \Omega_J
    \to 
    {S}^{j_1...j_{p_J} l}
     \times I^l_\epsilon 
    :=
    \left\{ 
        (x^\mu) \in {S}^{j_1...j_{p_J} l}_{x^l = c}\left| 
    c \in \left[-\fr{\epsilon}{2}, \fr{\epsilon}{2}\right]
    \right\}\right. \,.
\end{equation}
The delta function form for $\Omega_J$ can also be simplified in this limit as
\begin{equation}
\begin{split}
    \delta_{p_J } ({S}^{j_1...j_{p_J} l}
     \times I^l_\epsilon 
    )
 &
= 
\fr{\epsilon_{\mu_1... \mu_{D-p_J} \nu_1...\nu_{p_J}}}{(D-p_J)!p_J!}
\rmd x^{\nu_1} \weds \rmd  x^{\nu_{p_J}}
\int_{{S}^{j_1...j_{p_J} l}
     \times I^l_\epsilon } 
\rmd  z^{\mu_1}\weds \rmd z^{\mu_{D-p_J}}
\delta^D (x-z)    
 \\
 &
 = 
\delta(x^{j_1}) 
\cdots
\delta(x^{j_{p_J}}) 
\Theta \left(x^l; \left[-\fr{\epsilon}{2},\fr{\epsilon}{2}\right]\right)
\rmd x^{j_1} \weds \rmd x^{j_{p_J}}.
\end{split}
\end{equation}
Here, $\Theta \left(x^l; \left[-\fr{\epsilon}{2},\fr{\epsilon}{2}\right]\right)$ is a step function defined so that 
$\Theta \left(x^l; \left[-\fr{\epsilon}{2},\fr{\epsilon}{2}\right]\right) = 1 $
 if 
$x^l \in \left[-\fr{\epsilon}{2},\fr{\epsilon}{2}\right]$, 
and 
$\Theta \left(x^l; \left[-\fr{\epsilon}{2},\fr{\epsilon}{2}\right]\right) =0$ otherwise.
By this choice, the intersection $\Sigma_I \cap \Omega_J$ becomes
\begin{equation}
 \Sigma_I \cap \Omega_J 
\to S^{i_1...i_{p_I}l} \cap  S^{j_1...j_{p_J}l}
 = S^{i_1...i_{p_I}l j_1...j_{p_J}},
\end{equation}
and the delta function form in \er{correlation_RHS} can be simplified as 
\begin{equation}
\begin{split}
\delta_{p_I+1} (\Sigma_I)
\wed \delta_{p_J} (\Omega_J)
&\to \delta_{p_I+ p_J +1} (S^{i_1...i_{p_I}l j_1...j_{p_J}})
\\
& 
= 
\delta(x^{i_1}) 
\cdots
\delta(x^{i_{p_I}}) 
\delta(x^{l}) 
\delta(x^{j_1}) 
\cdots
\delta(x^{j_{p_J}}) 
\rmd x^{i_1} \weds \rmd  x^{i_{p_I}} \wed \rmd x^l
\wed 
\rmd x^{j_1} \weds \rmd  x^{j_{p_J}} \,.
\end{split}
\end{equation}

Now, we can evaluate the integral in \er{correlation_suppl}.
Since $\b{S}^{i_1...i_p l}_{x^l = \epsilon}$ does not intersect with $\Omega_J$, we have 
\begin{equation}
    \begin{split}
&        \int \rmd A_{IJ} \wed \delta_{p_I+1} 
(\Sigma_I)
\wed \delta_{p_J} (\Omega_J )
\to 
 \int \rmd A_{IJ} \wed
 \delta_{p_I+ p_J +1} (S^{i_1...i_{p_I}l j_1...j_{p_J}})
\\
&
=  
(-1)^{p_{IJ}(D-p_{IJ}-1)+1}
\int \rmd^D x
M_{IJ}^{i_1...i_{p_I} l j_1...j_{p_J}}
\delta(x^{i_1}) \cdots
\delta(x^{i_{p_I}})
 \delta(x^{l}) 
\delta(x^{j_1}) \cdots \delta(x^{j_{p_J}}) 
\\
&
=
(-1)^{p_{IJ}(D-p_{IJ}-1)+1}
M_{IJ}^{i_1...i_{p_I} l j_1...j_{p_J}}
\vol(S^{i_1...i_{p_I}l j_1...j_{p_J}}).
\end{split}
\end{equation}
Therefore, we arrive at 
\begin{equation}
    \fr{  \rmi \vevs{Q_I (\Sigma_I )
Q_J(\Sigma_J)}}{\vol (\Sigma_I \cap \Omega_J)}
 \to 
(-1)^{p_{IJ} (D-p_{IJ}-1)+1}
M_{IJ}^{i_1...i_{p_I} l j_1...j_{p_J}}\,.
\end{equation}

\begin{thebibliography}{47}%
\makeatletter
\providecommand \@ifxundefined [1]{%
 \@ifx{#1\undefined}
}%
\providecommand \@ifnum [1]{%
 \ifnum #1\expandafter \@firstoftwo
 \else \expandafter \@secondoftwo
 \fi
}%
\providecommand \@ifx [1]{%
 \ifx #1\expandafter \@firstoftwo
 \else \expandafter \@secondoftwo
 \fi
}%
\providecommand \natexlab [1]{#1}%
\providecommand \enquote  [1]{``#1''}%
\providecommand \bibnamefont  [1]{#1}%
\providecommand \bibfnamefont [1]{#1}%
\providecommand \citenamefont [1]{#1}%
\providecommand \href@noop [0]{\@secondoftwo}%
\providecommand \href [0]{\begingroup \@sanitize@url \@href}%
\providecommand \@href[1]{\@@startlink{#1}\@@href}%
\providecommand \@@href[1]{\endgroup#1\@@endlink}%
\providecommand \@sanitize@url [0]{\catcode `\\12\catcode `\$12\catcode
  `\&12\catcode `\#12\catcode `\^12\catcode `\_12\catcode `\%12\relax}%
\providecommand \@@startlink[1]{}%
\providecommand \@@endlink[0]{}%
\providecommand \url  [0]{\begingroup\@sanitize@url \@url }%
\providecommand \@url [1]{\endgroup\@href {#1}{\urlprefix }}%
\providecommand \urlprefix  [0]{URL }%
\providecommand \Eprint [0]{\href }%
\providecommand \doibase [0]{http://dx.doi.org/}%
\providecommand \selectlanguage [0]{\@gobble}%
\providecommand \bibinfo  [0]{\@secondoftwo}%
\providecommand \bibfield  [0]{\@secondoftwo}%
\providecommand \translation [1]{[#1]}%
\providecommand \BibitemOpen [0]{}%
\providecommand \bibitemStop [0]{}%
\providecommand \bibitemNoStop [0]{.\EOS\space}%
\providecommand \EOS [0]{\spacefactor3000\relax}%
\providecommand \BibitemShut  [1]{\csname bibitem#1\endcsname}%
\let\auto@bib@innerbib\@empty
\bibitem [{\citenamefont {Nambu}\ and\ \citenamefont
  {Jona-Lasinio}(1961)}]{Nambu:1961tp}%
  \BibitemOpen
  \bibfield  {author} {\bibinfo {author} {\bibfnamefont {Y.}~\bibnamefont
  {Nambu}}\ and\ \bibinfo {author} {\bibfnamefont {G.}~\bibnamefont
  {Jona-Lasinio}},\ }\href {\doibase 10.1103/PhysRev.122.345} {\bibfield
  {journal} {\bibinfo  {journal} {Phys. Rev.}\ }\textbf {\bibinfo {volume}
  {122}},\ \bibinfo {pages} {345} (\bibinfo {year} {1961})}\BibitemShut
  {NoStop}%
\bibitem [{\citenamefont {Goldstone}(1961)}]{Goldstone:1961eq}%
  \BibitemOpen
  \bibfield  {author} {\bibinfo {author} {\bibfnamefont {J.}~\bibnamefont
  {Goldstone}},\ }\href {\doibase 10.1007/BF02812722} {\bibfield  {journal}
  {\bibinfo  {journal} {Nuovo Cim.}\ }\textbf {\bibinfo {volume} {19}},\
  \bibinfo {pages} {154} (\bibinfo {year} {1961})}\BibitemShut {NoStop}%
\bibitem [{\citenamefont {Goldstone}\ \emph {et~al.}(1962)\citenamefont
  {Goldstone}, \citenamefont {Salam},\ and\ \citenamefont
  {Weinberg}}]{Goldstone:1962es}%
  \BibitemOpen
  \bibfield  {author} {\bibinfo {author} {\bibfnamefont {J.}~\bibnamefont
  {Goldstone}}, \bibinfo {author} {\bibfnamefont {A.}~\bibnamefont {Salam}}, \
  and\ \bibinfo {author} {\bibfnamefont {S.}~\bibnamefont {Weinberg}},\ }\href
  {\doibase 10.1103/PhysRev.127.965} {\bibfield  {journal} {\bibinfo  {journal}
  {Phys. Rev.}\ }\textbf {\bibinfo {volume} {127}},\ \bibinfo {pages} {965}
  (\bibinfo {year} {1962})}\BibitemShut {NoStop}%
\bibitem [{\citenamefont {Nielsen}\ and\ \citenamefont
  {Chadha}(1976)}]{Nielsen:1975hm}%
  \BibitemOpen
  \bibfield  {author} {\bibinfo {author} {\bibfnamefont {H.~B.}\ \bibnamefont
  {Nielsen}}\ and\ \bibinfo {author} {\bibfnamefont {S.}~\bibnamefont
  {Chadha}},\ }\href {\doibase 10.1016/0550-3213(76)90025-0} {\bibfield
  {journal} {\bibinfo  {journal} {Nucl. Phys.}\ }\textbf {\bibinfo {volume}
  {B105}},\ \bibinfo {pages} {445} (\bibinfo {year} {1976})}\BibitemShut
  {NoStop}%
\bibitem [{\citenamefont {Miransky}\ and\ \citenamefont
  {Shovkovy}(2002)}]{Miransky:2001tw}%
  \BibitemOpen
  \bibfield  {author} {\bibinfo {author} {\bibfnamefont {V.~A.}\ \bibnamefont
  {Miransky}}\ and\ \bibinfo {author} {\bibfnamefont {I.~A.}\ \bibnamefont
  {Shovkovy}},\ }\href {\doibase 10.1103/PhysRevLett.88.111601} {\bibfield
  {journal} {\bibinfo  {journal} {Phys. Rev. Lett.}\ }\textbf {\bibinfo
  {volume} {88}},\ \bibinfo {pages} {111601} (\bibinfo {year} {2002})},\
  \Eprint {http://arxiv.org/abs/hep-ph/0108178} {arXiv:hep-ph/0108178 [hep-ph]}
  \BibitemShut {NoStop}%
\bibitem [{\citenamefont {Sch\"afer}\ \emph {et~al.}(2001)\citenamefont
  {Sch\"afer}, \citenamefont {Son}, \citenamefont {Stephanov}, \citenamefont
  {Toublan},\ and\ \citenamefont {Verbaarschot}}]{Schafer:2001bq}%
  \BibitemOpen
  \bibfield  {author} {\bibinfo {author} {\bibfnamefont {T.}~\bibnamefont
  {Sch\"afer}}, \bibinfo {author} {\bibfnamefont {D.~T.}\ \bibnamefont {Son}},
  \bibinfo {author} {\bibfnamefont {M.~A.}\ \bibnamefont {Stephanov}}, \bibinfo
  {author} {\bibfnamefont {D.}~\bibnamefont {Toublan}}, \ and\ \bibinfo
  {author} {\bibfnamefont {J.~J.~M.}\ \bibnamefont {Verbaarschot}},\ }\href
  {\doibase 10.1016/S0370-2693(01)01265-5} {\bibfield  {journal} {\bibinfo
  {journal} {Phys. Lett. B}\ }\textbf {\bibinfo {volume} {522}},\ \bibinfo
  {pages} {67} (\bibinfo {year} {2001})},\ \Eprint
  {http://arxiv.org/abs/hep-ph/0108210} {arXiv:hep-ph/0108210} \BibitemShut
  {NoStop}%
\bibitem [{\citenamefont {Nambu}(2004)}]{Nambu:2004yia}%
  \BibitemOpen
  \bibfield  {author} {\bibinfo {author} {\bibfnamefont {Y.}~\bibnamefont
  {Nambu}},\ }\href {\doibase 10.1023/B:JOSS.0000019827.74407.2d} {\bibfield
  {journal} {\bibinfo  {journal} {J. Statist. Phys.}\ }\textbf {\bibinfo
  {volume} {115}},\ \bibinfo {pages} {7} (\bibinfo {year} {2004})}\BibitemShut
  {NoStop}%
\bibitem [{\citenamefont {Watanabe}\ and\ \citenamefont
  {Brauner}(2011)}]{Watanabe:2011ec}%
  \BibitemOpen
  \bibfield  {author} {\bibinfo {author} {\bibfnamefont {H.}~\bibnamefont
  {Watanabe}}\ and\ \bibinfo {author} {\bibfnamefont {T.}~\bibnamefont
  {Brauner}},\ }\href {\doibase 10.1103/PhysRevD.84.125013} {\bibfield
  {journal} {\bibinfo  {journal} {Phys. Rev. D}\ }\textbf {\bibinfo {volume}
  {84}},\ \bibinfo {pages} {125013} (\bibinfo {year} {2011})},\ \Eprint
  {http://arxiv.org/abs/1109.6327} {arXiv:1109.6327 [hep-ph]} \BibitemShut
  {NoStop}%
\bibitem [{\citenamefont {Watanabe}\ and\ \citenamefont
  {Murayama}(2012)}]{Watanabe:2012hr}%
  \BibitemOpen
  \bibfield  {author} {\bibinfo {author} {\bibfnamefont {H.}~\bibnamefont
  {Watanabe}}\ and\ \bibinfo {author} {\bibfnamefont {H.}~\bibnamefont
  {Murayama}},\ }\href {\doibase 10.1103/PhysRevLett.108.251602} {\bibfield
  {journal} {\bibinfo  {journal} {Phys. Rev. Lett.}\ }\textbf {\bibinfo
  {volume} {108}},\ \bibinfo {pages} {251602} (\bibinfo {year} {2012})},\
  \Eprint {http://arxiv.org/abs/1203.0609} {arXiv:1203.0609 [hep-th]}
  \BibitemShut {NoStop}%
\bibitem [{\citenamefont {Hidaka}(2013)}]{Hidaka:2012ym}%
  \BibitemOpen
  \bibfield  {author} {\bibinfo {author} {\bibfnamefont {Y.}~\bibnamefont
  {Hidaka}},\ }\href {\doibase 10.1103/PhysRevLett.110.091601} {\bibfield
  {journal} {\bibinfo  {journal} {Phys. Rev. Lett.}\ }\textbf {\bibinfo
  {volume} {110}},\ \bibinfo {pages} {091601} (\bibinfo {year} {2013})},\
  \Eprint {http://arxiv.org/abs/1203.1494} {arXiv:1203.1494 [hep-th]}
  \BibitemShut {NoStop}%
\bibitem [{\citenamefont {Watanabe}\ and\ \citenamefont
  {Murayama}(2014)}]{Watanabe:2014fva}%
  \BibitemOpen
  \bibfield  {author} {\bibinfo {author} {\bibfnamefont {H.}~\bibnamefont
  {Watanabe}}\ and\ \bibinfo {author} {\bibfnamefont {H.}~\bibnamefont
  {Murayama}},\ }\href {\doibase 10.1103/PhysRevX.4.031057} {\bibfield
  {journal} {\bibinfo  {journal} {Phys. Rev.}\ }\textbf {\bibinfo {volume}
  {X4}},\ \bibinfo {pages} {031057} (\bibinfo {year} {2014})},\ \Eprint
  {http://arxiv.org/abs/1402.7066} {arXiv:1402.7066 [hep-th]} \BibitemShut
  {NoStop}%
\bibitem [{\citenamefont {Gaiotto}\ \emph {et~al.}(2015)\citenamefont
  {Gaiotto}, \citenamefont {Kapustin}, \citenamefont {Seiberg},\ and\
  \citenamefont {Willett}}]{Gaiotto:2014kfa}%
  \BibitemOpen
  \bibfield  {author} {\bibinfo {author} {\bibfnamefont {D.}~\bibnamefont
  {Gaiotto}}, \bibinfo {author} {\bibfnamefont {A.}~\bibnamefont {Kapustin}},
  \bibinfo {author} {\bibfnamefont {N.}~\bibnamefont {Seiberg}}, \ and\
  \bibinfo {author} {\bibfnamefont {B.}~\bibnamefont {Willett}},\ }\href
  {\doibase 10.1007/JHEP02(2015)172} {\bibfield  {journal} {\bibinfo  {journal}
  {JHEP}\ }\textbf {\bibinfo {volume} {02}},\ \bibinfo {pages} {172} (\bibinfo
  {year} {2015})},\ \Eprint {http://arxiv.org/abs/1412.5148} {arXiv:1412.5148
  [hep-th]} \BibitemShut {NoStop}%
\bibitem [{\citenamefont {Batista}\ and\ \citenamefont
  {Nussinov}(2005)}]{Batista:2004sc}%
  \BibitemOpen
  \bibfield  {author} {\bibinfo {author} {\bibfnamefont {C.~D.}\ \bibnamefont
  {Batista}}\ and\ \bibinfo {author} {\bibfnamefont {Z.}~\bibnamefont
  {Nussinov}},\ }\href {\doibase 10.1103/PhysRevB.72.045137} {\bibfield
  {journal} {\bibinfo  {journal} {Phys. Rev. B}\ }\textbf {\bibinfo {volume}
  {72}},\ \bibinfo {pages} {045137} (\bibinfo {year} {2005})},\ \Eprint
  {http://arxiv.org/abs/cond-mat/0410599} {arXiv:cond-mat/0410599} \BibitemShut
  {NoStop}%
\bibitem [{\citenamefont {Pantev}\ and\ \citenamefont
  {Sharpe}(2006{\natexlab{a}})}]{Pantev:2005zs}%
  \BibitemOpen
  \bibfield  {author} {\bibinfo {author} {\bibfnamefont {T.}~\bibnamefont
  {Pantev}}\ and\ \bibinfo {author} {\bibfnamefont {E.}~\bibnamefont
  {Sharpe}},\ }\href {\doibase 10.4310/ATMP.2006.v10.n1.a4} {\bibfield
  {journal} {\bibinfo  {journal} {Adv. Theor. Math. Phys.}\ }\textbf {\bibinfo
  {volume} {10}},\ \bibinfo {pages} {77} (\bibinfo {year}
  {2006}{\natexlab{a}})},\ \Eprint {http://arxiv.org/abs/hep-th/0502053}
  {arXiv:hep-th/0502053 [hep-th]} \BibitemShut {NoStop}%
\bibitem [{\citenamefont {Pantev}\ and\ \citenamefont
  {Sharpe}(2006{\natexlab{b}})}]{Pantev:2005wj}%
  \BibitemOpen
  \bibfield  {author} {\bibinfo {author} {\bibfnamefont {T.}~\bibnamefont
  {Pantev}}\ and\ \bibinfo {author} {\bibfnamefont {E.}~\bibnamefont
  {Sharpe}},\ }\href {\doibase 10.1016/j.nuclphysb.2005.10.035} {\bibfield
  {journal} {\bibinfo  {journal} {Nucl. Phys.}\ }\textbf {\bibinfo {volume}
  {B733}},\ \bibinfo {pages} {233} (\bibinfo {year} {2006}{\natexlab{b}})},\
  \Eprint {http://arxiv.org/abs/hep-th/0502044} {arXiv:hep-th/0502044 [hep-th]}
  \BibitemShut {NoStop}%
\bibitem [{\citenamefont {Pantev}\ and\ \citenamefont
  {Sharpe}(2005)}]{Pantev:2005rh}%
  \BibitemOpen
  \bibfield  {author} {\bibinfo {author} {\bibfnamefont {T.}~\bibnamefont
  {Pantev}}\ and\ \bibinfo {author} {\bibfnamefont {E.}~\bibnamefont
  {Sharpe}},\ }\href@noop {} {\  (\bibinfo {year} {2005})},\ \Eprint
  {http://arxiv.org/abs/hep-th/0502027} {arXiv:hep-th/0502027 [hep-th]}
  \BibitemShut {NoStop}%
\bibitem [{\citenamefont {Nussinov}\ and\ \citenamefont
  {Ortiz}(2009{\natexlab{a}})}]{Nussinov:2006iva}%
  \BibitemOpen
  \bibfield  {author} {\bibinfo {author} {\bibfnamefont {Z.}~\bibnamefont
  {Nussinov}}\ and\ \bibinfo {author} {\bibfnamefont {G.}~\bibnamefont
  {Ortiz}},\ }\href {\doibase 10.1073/pnas.0803726105} {\bibfield  {journal}
  {\bibinfo  {journal} {Proc. Nat. Acad. Sci.}\ }\textbf {\bibinfo {volume}
  {106}},\ \bibinfo {pages} {16944} (\bibinfo {year} {2009}{\natexlab{a}})},\
  \Eprint {http://arxiv.org/abs/cond-mat/0605316} {arXiv:cond-mat/0605316}
  \BibitemShut {NoStop}%
\bibitem [{\citenamefont {Nussinov}\ and\ \citenamefont
  {Ortiz}(2008)}]{Nussinov:2008aa}%
  \BibitemOpen
  \bibfield  {author} {\bibinfo {author} {\bibfnamefont {Z.}~\bibnamefont
  {Nussinov}}\ and\ \bibinfo {author} {\bibfnamefont {G.}~\bibnamefont
  {Ortiz}},\ }\href {\doibase 10.1103/physrevb.77.064302} {\bibfield  {journal}
  {\bibinfo  {journal} {Phys. Rev. B}\ }\textbf {\bibinfo {volume} {77}},\
  \bibinfo {pages} {064302} (\bibinfo {year} {2008})},\ \Eprint
  {http://arxiv.org/abs/0709.2717} {arXiv:0709.2717} \BibitemShut {NoStop}%
\bibitem [{\citenamefont {Nussinov}\ and\ \citenamefont
  {Ortiz}(2009{\natexlab{b}})}]{Nussinov:2009zz}%
  \BibitemOpen
  \bibfield  {author} {\bibinfo {author} {\bibfnamefont {Z.}~\bibnamefont
  {Nussinov}}\ and\ \bibinfo {author} {\bibfnamefont {G.}~\bibnamefont
  {Ortiz}},\ }\href {\doibase 10.1016/j.aop.2008.11.002} {\bibfield  {journal}
  {\bibinfo  {journal} {Annals Phys.}\ }\textbf {\bibinfo {volume} {324}},\
  \bibinfo {pages} {977} (\bibinfo {year} {2009}{\natexlab{b}})},\ \Eprint
  {http://arxiv.org/abs/cond-mat/0702377} {arXiv:cond-mat/0702377} \BibitemShut
  {NoStop}%
\bibitem [{\citenamefont {Nussinov}\ \emph {et~al.}(2012)\citenamefont
  {Nussinov}, \citenamefont {Ortiz},\ and\ \citenamefont
  {Cobanera}}]{Nussinov:2011mz}%
  \BibitemOpen
  \bibfield  {author} {\bibinfo {author} {\bibfnamefont {Z.}~\bibnamefont
  {Nussinov}}, \bibinfo {author} {\bibfnamefont {G.}~\bibnamefont {Ortiz}}, \
  and\ \bibinfo {author} {\bibfnamefont {E.}~\bibnamefont {Cobanera}},\ }\href
  {\doibase 10.1016/j.aop.2012.07.001} {\bibfield  {journal} {\bibinfo
  {journal} {Annals Phys.}\ }\textbf {\bibinfo {volume} {327}},\ \bibinfo
  {pages} {2491} (\bibinfo {year} {2012})},\ \Eprint
  {http://arxiv.org/abs/1110.2179} {arXiv:1110.2179 [cond-mat.stat-mech]}
  \BibitemShut {NoStop}%
\bibitem [{\citenamefont {Banks}\ and\ \citenamefont
  {Seiberg}(2011)}]{Banks:2010zn}%
  \BibitemOpen
  \bibfield  {author} {\bibinfo {author} {\bibfnamefont {T.}~\bibnamefont
  {Banks}}\ and\ \bibinfo {author} {\bibfnamefont {N.}~\bibnamefont
  {Seiberg}},\ }\href {\doibase 10.1103/PhysRevD.83.084019} {\bibfield
  {journal} {\bibinfo  {journal} {Phys. Rev.}\ }\textbf {\bibinfo {volume}
  {D83}},\ \bibinfo {pages} {084019} (\bibinfo {year} {2011})},\ \Eprint
  {http://arxiv.org/abs/1011.5120} {arXiv:1011.5120 [hep-th]} \BibitemShut
  {NoStop}%
\bibitem [{\citenamefont {Distler}\ and\ \citenamefont
  {Sharpe}(2011)}]{Distler:2010zg}%
  \BibitemOpen
  \bibfield  {author} {\bibinfo {author} {\bibfnamefont {J.}~\bibnamefont
  {Distler}}\ and\ \bibinfo {author} {\bibfnamefont {E.}~\bibnamefont
  {Sharpe}},\ }\href {\doibase 10.1103/PhysRevD.83.085010} {\bibfield
  {journal} {\bibinfo  {journal} {Phys. Rev.}\ }\textbf {\bibinfo {volume}
  {D83}},\ \bibinfo {pages} {085010} (\bibinfo {year} {2011})},\ \Eprint
  {http://arxiv.org/abs/1008.0419} {arXiv:1008.0419 [hep-th]} \BibitemShut
  {NoStop}%
\bibitem [{\citenamefont {Kapustin}\ and\ \citenamefont
  {Seiberg}(2014)}]{Kapustin:2014gua}%
  \BibitemOpen
  \bibfield  {author} {\bibinfo {author} {\bibfnamefont {A.}~\bibnamefont
  {Kapustin}}\ and\ \bibinfo {author} {\bibfnamefont {N.}~\bibnamefont
  {Seiberg}},\ }\href {\doibase 10.1007/JHEP04(2014)001} {\bibfield  {journal}
  {\bibinfo  {journal} {JHEP}\ }\textbf {\bibinfo {volume} {04}},\ \bibinfo
  {pages} {001} (\bibinfo {year} {2014})},\ \Eprint
  {http://arxiv.org/abs/1401.0740} {arXiv:1401.0740 [hep-th]} \BibitemShut
  {NoStop}%
\bibitem [{\citenamefont {Kugo}\ \emph {et~al.}(1985)\citenamefont {Kugo},
  \citenamefont {Terao},\ and\ \citenamefont {Uehara}}]{Kugo:1985jc}%
  \BibitemOpen
  \bibfield  {author} {\bibinfo {author} {\bibfnamefont {T.}~\bibnamefont
  {Kugo}}, \bibinfo {author} {\bibfnamefont {H.}~\bibnamefont {Terao}}, \ and\
  \bibinfo {author} {\bibfnamefont {S.}~\bibnamefont {Uehara}},\ }\href@noop {}
  {\bibfield  {journal} {\bibinfo  {journal} {Prog. Theor. Phys. Suppl.}\
  }\textbf {\bibinfo {volume} {85}},\ \bibinfo {pages} {122} (\bibinfo {year}
  {1985})}\BibitemShut {NoStop}%
\bibitem [{\citenamefont {Kovner}\ and\ \citenamefont
  {Rosenstein}(1994)}]{Kovner:1992pu}%
  \BibitemOpen
  \bibfield  {author} {\bibinfo {author} {\bibfnamefont {A.}~\bibnamefont
  {Kovner}}\ and\ \bibinfo {author} {\bibfnamefont {B.}~\bibnamefont
  {Rosenstein}},\ }\href {\doibase 10.1103/PhysRevD.49.5571} {\bibfield
  {journal} {\bibinfo  {journal} {Phys. Rev. D}\ }\textbf {\bibinfo {volume}
  {49}},\ \bibinfo {pages} {5571} (\bibinfo {year} {1994})},\ \Eprint
  {http://arxiv.org/abs/hep-th/9210154} {arXiv:hep-th/9210154} \BibitemShut
  {NoStop}%
\bibitem [{\citenamefont {Lake}(2018)}]{Lake:2018dqm}%
  \BibitemOpen
  \bibfield  {author} {\bibinfo {author} {\bibfnamefont {E.}~\bibnamefont
  {Lake}},\ }\href@noop {} {\  (\bibinfo {year} {2018})},\ \Eprint
  {http://arxiv.org/abs/1802.07747} {arXiv:1802.07747 [hep-th]} \BibitemShut
  {NoStop}%
\bibitem [{\citenamefont {Hidaka}\ \emph
  {et~al.}(2021{\natexlab{a}})\citenamefont {Hidaka}, \citenamefont {Hirono},\
  and\ \citenamefont {Yokokura}}]{Hidaka:2020ucc}%
  \BibitemOpen
  \bibfield  {author} {\bibinfo {author} {\bibfnamefont {Y.}~\bibnamefont
  {Hidaka}}, \bibinfo {author} {\bibfnamefont {Y.}~\bibnamefont {Hirono}}, \
  and\ \bibinfo {author} {\bibfnamefont {R.}~\bibnamefont {Yokokura}},\ }\href
  {\doibase 10.1103/PhysRevLett.126.071601} {\bibfield  {journal} {\bibinfo
  {journal} {Phys. Rev. Lett.}\ }\textbf {\bibinfo {volume} {126}},\ \bibinfo
  {pages} {071601} (\bibinfo {year} {2021}{\natexlab{a}})},\ \Eprint
  {http://arxiv.org/abs/2007.15901} {arXiv:2007.15901 [hep-th]} \BibitemShut
  {NoStop}%
\bibitem [{\citenamefont {Carroll}\ \emph {et~al.}(1990)\citenamefont
  {Carroll}, \citenamefont {Field},\ and\ \citenamefont
  {Jackiw}}]{Carroll:1989vb}%
  \BibitemOpen
  \bibfield  {author} {\bibinfo {author} {\bibfnamefont {S.~M.}\ \bibnamefont
  {Carroll}}, \bibinfo {author} {\bibfnamefont {G.~B.}\ \bibnamefont {Field}},
  \ and\ \bibinfo {author} {\bibfnamefont {R.}~\bibnamefont {Jackiw}},\ }\href
  {\doibase 10.1103/PhysRevD.41.1231} {\bibfield  {journal} {\bibinfo
  {journal} {Phys. Rev. D}\ }\textbf {\bibinfo {volume} {41}},\ \bibinfo
  {pages} {1231} (\bibinfo {year} {1990})}\BibitemShut {NoStop}%
\bibitem [{\citenamefont {Garretson}\ \emph {et~al.}(1992)\citenamefont
  {Garretson}, \citenamefont {Field},\ and\ \citenamefont
  {Carroll}}]{Garretson:1992vt}%
  \BibitemOpen
  \bibfield  {author} {\bibinfo {author} {\bibfnamefont {W.~D.}\ \bibnamefont
  {Garretson}}, \bibinfo {author} {\bibfnamefont {G.~B.}\ \bibnamefont
  {Field}}, \ and\ \bibinfo {author} {\bibfnamefont {S.~M.}\ \bibnamefont
  {Carroll}},\ }\href {\doibase 10.1103/PhysRevD.46.5346} {\bibfield  {journal}
  {\bibinfo  {journal} {Phys. Rev. D}\ }\textbf {\bibinfo {volume} {46}},\
  \bibinfo {pages} {5346} (\bibinfo {year} {1992})},\ \Eprint
  {http://arxiv.org/abs/hep-ph/9209238} {arXiv:hep-ph/9209238} \BibitemShut
  {NoStop}%
\bibitem [{\citenamefont {Anber}\ and\ \citenamefont
  {Sorbo}(2006)}]{Anber:2006xt}%
  \BibitemOpen
  \bibfield  {author} {\bibinfo {author} {\bibfnamefont {M.~M.}\ \bibnamefont
  {Anber}}\ and\ \bibinfo {author} {\bibfnamefont {L.}~\bibnamefont {Sorbo}},\
  }\href {\doibase 10.1088/1475-7516/2006/10/018} {\bibfield  {journal}
  {\bibinfo  {journal} {JCAP}\ }\textbf {\bibinfo {volume} {0610}},\ \bibinfo
  {pages} {018} (\bibinfo {year} {2006})},\ \Eprint
  {http://arxiv.org/abs/astro-ph/0606534} {arXiv:astro-ph/0606534 [astro-ph]}
  \BibitemShut {NoStop}%
\bibitem [{\citenamefont {Joyce}\ and\ \citenamefont
  {Shaposhnikov}(1997)}]{Joyce:1997uy}%
  \BibitemOpen
  \bibfield  {author} {\bibinfo {author} {\bibfnamefont {M.}~\bibnamefont
  {Joyce}}\ and\ \bibinfo {author} {\bibfnamefont {M.~E.}\ \bibnamefont
  {Shaposhnikov}},\ }\href {\doibase 10.1103/PhysRevLett.79.1193} {\bibfield
  {journal} {\bibinfo  {journal} {Phys. Rev. Lett.}\ }\textbf {\bibinfo
  {volume} {79}},\ \bibinfo {pages} {1193} (\bibinfo {year} {1997})},\ \Eprint
  {http://arxiv.org/abs/astro-ph/9703005} {arXiv:astro-ph/9703005} \BibitemShut
  {NoStop}%
\bibitem [{\citenamefont {Akamatsu}\ and\ \citenamefont
  {Yamamoto}(2013)}]{Akamatsu:2013pjd}%
  \BibitemOpen
  \bibfield  {author} {\bibinfo {author} {\bibfnamefont {Y.}~\bibnamefont
  {Akamatsu}}\ and\ \bibinfo {author} {\bibfnamefont {N.}~\bibnamefont
  {Yamamoto}},\ }\href {\doibase 10.1103/PhysRevLett.111.052002} {\bibfield
  {journal} {\bibinfo  {journal} {Phys. Rev. Lett.}\ }\textbf {\bibinfo
  {volume} {111}},\ \bibinfo {pages} {052002} (\bibinfo {year} {2013})},\
  \Eprint {http://arxiv.org/abs/1302.2125} {arXiv:1302.2125 [nucl-th]}
  \BibitemShut {NoStop}%
\bibitem [{\citenamefont {Ooguri}\ and\ \citenamefont
  {Oshikawa}(2012)}]{Ooguri:2011aa}%
  \BibitemOpen
  \bibfield  {author} {\bibinfo {author} {\bibfnamefont {H.}~\bibnamefont
  {Ooguri}}\ and\ \bibinfo {author} {\bibfnamefont {M.}~\bibnamefont
  {Oshikawa}},\ }\href {\doibase 10.1103/PhysRevLett.108.161803} {\bibfield
  {journal} {\bibinfo  {journal} {Phys. Rev. Lett.}\ }\textbf {\bibinfo
  {volume} {108}},\ \bibinfo {pages} {161803} (\bibinfo {year} {2012})},\
  \Eprint {http://arxiv.org/abs/1112.1414} {arXiv:1112.1414
  [cond-mat.mes-hall]} \BibitemShut {NoStop}%
\bibitem [{\citenamefont {Nakamura}\ \emph {et~al.}(2010)\citenamefont
  {Nakamura}, \citenamefont {Ooguri},\ and\ \citenamefont
  {Park}}]{Nakamura:2009tf}%
  \BibitemOpen
  \bibfield  {author} {\bibinfo {author} {\bibfnamefont {S.}~\bibnamefont
  {Nakamura}}, \bibinfo {author} {\bibfnamefont {H.}~\bibnamefont {Ooguri}}, \
  and\ \bibinfo {author} {\bibfnamefont {C.-S.}\ \bibnamefont {Park}},\ }\href
  {\doibase 10.1103/PhysRevD.81.044018} {\bibfield  {journal} {\bibinfo
  {journal} {Phys. Rev.}\ }\textbf {\bibinfo {volume} {D81}},\ \bibinfo {pages}
  {044018} (\bibinfo {year} {2010})},\ \Eprint {http://arxiv.org/abs/0911.0679}
  {arXiv:0911.0679 [hep-th]} \BibitemShut {NoStop}%
\bibitem [{\citenamefont {Yamamoto}(2016)}]{Yamamoto:2015maz}%
  \BibitemOpen
  \bibfield  {author} {\bibinfo {author} {\bibfnamefont {N.}~\bibnamefont
  {Yamamoto}},\ }\href {\doibase 10.1103/PhysRevD.93.085036} {\bibfield
  {journal} {\bibinfo  {journal} {Phys. Rev.}\ }\textbf {\bibinfo {volume}
  {D93}},\ \bibinfo {pages} {085036} (\bibinfo {year} {2016})},\ \Eprint
  {http://arxiv.org/abs/1512.05668} {arXiv:1512.05668 [hep-th]} \BibitemShut
  {NoStop}%
\bibitem [{\citenamefont {Sogabe}\ and\ \citenamefont
  {Yamamoto}(2019)}]{Sogabe:2019gif}%
  \BibitemOpen
  \bibfield  {author} {\bibinfo {author} {\bibfnamefont {N.}~\bibnamefont
  {Sogabe}}\ and\ \bibinfo {author} {\bibfnamefont {N.}~\bibnamefont
  {Yamamoto}},\ }\href {\doibase 10.1103/PhysRevD.99.125003} {\bibfield
  {journal} {\bibinfo  {journal} {Phys. Rev. D}\ }\textbf {\bibinfo {volume}
  {99}},\ \bibinfo {pages} {125003} (\bibinfo {year} {2019})},\ \Eprint
  {http://arxiv.org/abs/1903.02846} {arXiv:1903.02846 [hep-th]} \BibitemShut
  {NoStop}%
\bibitem [{\citenamefont {Brauner}(2021)}]{Brauner:2020rtz}%
  \BibitemOpen
  \bibfield  {author} {\bibinfo {author} {\bibfnamefont {T.}~\bibnamefont
  {Brauner}},\ }\href {\doibase 10.1007/JHEP04(2021)045} {\bibfield  {journal}
  {\bibinfo  {journal} {JHEP}\ }\textbf {\bibinfo {volume} {04}},\ \bibinfo
  {pages} {045} (\bibinfo {year} {2021})},\ \Eprint
  {http://arxiv.org/abs/2012.00051} {arXiv:2012.00051 [hep-th]} \BibitemShut
  {NoStop}%
\bibitem [{\citenamefont {Heidenreich}\ \emph {et~al.}(2021)\citenamefont
  {Heidenreich}, \citenamefont {McNamara}, \citenamefont {Montero},
  \citenamefont {Reece}, \citenamefont {Rudelius},\ and\ \citenamefont
  {Valenzuela}}]{Heidenreich:2020pkc}%
  \BibitemOpen
  \bibfield  {author} {\bibinfo {author} {\bibfnamefont {B.}~\bibnamefont
  {Heidenreich}}, \bibinfo {author} {\bibfnamefont {J.}~\bibnamefont
  {McNamara}}, \bibinfo {author} {\bibfnamefont {M.}~\bibnamefont {Montero}},
  \bibinfo {author} {\bibfnamefont {M.}~\bibnamefont {Reece}}, \bibinfo
  {author} {\bibfnamefont {T.}~\bibnamefont {Rudelius}}, \ and\ \bibinfo
  {author} {\bibfnamefont {I.}~\bibnamefont {Valenzuela}},\ }\href {\doibase
  10.1007/JHEP11(2021)053} {\bibfield  {journal} {\bibinfo  {journal} {JHEP}\
  }\textbf {\bibinfo {volume} {11}},\ \bibinfo {pages} {053} (\bibinfo {year}
  {2021})},\ \Eprint {http://arxiv.org/abs/2012.00009} {arXiv:2012.00009
  [hep-th]} \BibitemShut {NoStop}%
\bibitem [{\citenamefont {Brauner}\ and\ \citenamefont
  {Kadam}(2017)}]{Brauner:2017mui}%
  \BibitemOpen
  \bibfield  {author} {\bibinfo {author} {\bibfnamefont {T.}~\bibnamefont
  {Brauner}}\ and\ \bibinfo {author} {\bibfnamefont {S.}~\bibnamefont
  {Kadam}},\ }\href {\doibase 10.1007/JHEP03(2017)015} {\bibfield  {journal}
  {\bibinfo  {journal} {JHEP}\ }\textbf {\bibinfo {volume} {03}},\ \bibinfo
  {pages} {015} (\bibinfo {year} {2017})},\ \Eprint
  {http://arxiv.org/abs/1701.06793} {arXiv:1701.06793 [hep-ph]} \BibitemShut
  {NoStop}%
\bibitem [{\citenamefont {Redlich}\ and\ \citenamefont
  {Wijewardhana}(1985)}]{Redlich:1984md}%
  \BibitemOpen
  \bibfield  {author} {\bibinfo {author} {\bibfnamefont {A.~N.}\ \bibnamefont
  {Redlich}}\ and\ \bibinfo {author} {\bibfnamefont {L.~C.~R.}\ \bibnamefont
  {Wijewardhana}},\ }\href {\doibase 10.1103/PhysRevLett.54.970} {\bibfield
  {journal} {\bibinfo  {journal} {Phys. Rev. Lett.}\ }\textbf {\bibinfo
  {volume} {54}},\ \bibinfo {pages} {970} (\bibinfo {year} {1985})}\BibitemShut
  {NoStop}%
\bibitem [{\citenamefont {Vilenkin}(1980)}]{Vilenkin:1980fu}%
  \BibitemOpen
  \bibfield  {author} {\bibinfo {author} {\bibfnamefont {A.}~\bibnamefont
  {Vilenkin}},\ }\href {\doibase 10.1103/PhysRevD.22.3080} {\bibfield
  {journal} {\bibinfo  {journal} {Phys. Rev. D}\ }\textbf {\bibinfo {volume}
  {22}},\ \bibinfo {pages} {3080} (\bibinfo {year} {1980})}\BibitemShut
  {NoStop}%
\bibitem [{\citenamefont {Nielsen}\ and\ \citenamefont
  {Ninomiya}(1983)}]{Nielsen:1983rb}%
  \BibitemOpen
  \bibfield  {author} {\bibinfo {author} {\bibfnamefont {H.~B.}\ \bibnamefont
  {Nielsen}}\ and\ \bibinfo {author} {\bibfnamefont {M.}~\bibnamefont
  {Ninomiya}},\ }\href {\doibase 10.1016/0370-2693(83)91529-0} {\bibfield
  {journal} {\bibinfo  {journal} {Phys. Lett. B}\ }\textbf {\bibinfo {volume}
  {130}},\ \bibinfo {pages} {389} (\bibinfo {year} {1983})}\BibitemShut
  {NoStop}%
\bibitem [{\citenamefont {Fukushima}\ \emph {et~al.}(2008)\citenamefont
  {Fukushima}, \citenamefont {Kharzeev},\ and\ \citenamefont
  {Warringa}}]{Fukushima:2008xe}%
  \BibitemOpen
  \bibfield  {author} {\bibinfo {author} {\bibfnamefont {K.}~\bibnamefont
  {Fukushima}}, \bibinfo {author} {\bibfnamefont {D.~E.}\ \bibnamefont
  {Kharzeev}}, \ and\ \bibinfo {author} {\bibfnamefont {H.~J.}\ \bibnamefont
  {Warringa}},\ }\href {\doibase 10.1103/PhysRevD.78.074033} {\bibfield
  {journal} {\bibinfo  {journal} {Phys. Rev. D}\ }\textbf {\bibinfo {volume}
  {78}},\ \bibinfo {pages} {074033} (\bibinfo {year} {2008})},\ \Eprint
  {http://arxiv.org/abs/0808.3382} {arXiv:0808.3382 [hep-ph]} \BibitemShut
  {NoStop}%
\bibitem [{\citenamefont {Minami}\ and\ \citenamefont
  {Hidaka}(2018)}]{Minami:2015uzo}%
  \BibitemOpen
  \bibfield  {author} {\bibinfo {author} {\bibfnamefont {Y.}~\bibnamefont
  {Minami}}\ and\ \bibinfo {author} {\bibfnamefont {Y.}~\bibnamefont
  {Hidaka}},\ }\href {\doibase 10.1103/PhysRevE.97.012130} {\bibfield
  {journal} {\bibinfo  {journal} {Phys. Rev. E}\ }\textbf {\bibinfo {volume}
  {97}},\ \bibinfo {pages} {012130} (\bibinfo {year} {2018})},\ \Eprint
  {http://arxiv.org/abs/1509.05042} {arXiv:1509.05042 [cond-mat.stat-mech]}
  \BibitemShut {NoStop}%
\bibitem [{\citenamefont {Cherman}\ \emph {et~al.}(2022)\citenamefont
  {Cherman}, \citenamefont {Jacobson},\ and\ \citenamefont
  {Neuzil}}]{Cherman:2021nox}%
  \BibitemOpen
  \bibfield  {author} {\bibinfo {author} {\bibfnamefont {A.}~\bibnamefont
  {Cherman}}, \bibinfo {author} {\bibfnamefont {T.}~\bibnamefont {Jacobson}}, \
  and\ \bibinfo {author} {\bibfnamefont {M.}~\bibnamefont {Neuzil}},\ }\href
  {\doibase 10.21468/SciPostPhys.12.4.116} {\bibfield  {journal} {\bibinfo
  {journal} {SciPost Phys.}\ }\textbf {\bibinfo {volume} {12}},\ \bibinfo
  {pages} {116} (\bibinfo {year} {2022})},\ \Eprint
  {http://arxiv.org/abs/2111.00078} {arXiv:2111.00078 [hep-th]} \BibitemShut
  {NoStop}%
\bibitem [{\citenamefont {Chen}\ \emph {et~al.}(2016)\citenamefont {Chen},
  \citenamefont {Tiwari},\ and\ \citenamefont {Ryu}}]{Chen:2015gma}%
  \BibitemOpen
  \bibfield  {author} {\bibinfo {author} {\bibfnamefont {X.}~\bibnamefont
  {Chen}}, \bibinfo {author} {\bibfnamefont {A.}~\bibnamefont {Tiwari}}, \ and\
  \bibinfo {author} {\bibfnamefont {S.}~\bibnamefont {Ryu}},\ }\href {\doibase
  10.1103/PhysRevB.94.045113} {\bibfield  {journal} {\bibinfo  {journal} {Phys.
  Rev.}\ }\textbf {\bibinfo {volume} {B94}},\ \bibinfo {pages} {045113}
  (\bibinfo {year} {2016})},\ \bibinfo {note} {[Addendum:
  \href{https://doi.org/10.1103/PhysRevB.94.079903}{{Phys. Rev.} {\bf B94}
  (2016) no.~7, 079903}]},\ \Eprint {http://arxiv.org/abs/1509.04266}
  {arXiv:1509.04266 [cond-mat.str-el]} \BibitemShut {NoStop}%
\bibitem [{\citenamefont {Hidaka}\ \emph
  {et~al.}(2021{\natexlab{b}})\citenamefont {Hidaka}, \citenamefont {Nitta},\
  and\ \citenamefont {Yokokura}}]{Hidaka:2020izy}%
  \BibitemOpen
  \bibfield  {author} {\bibinfo {author} {\bibfnamefont {Y.}~\bibnamefont
  {Hidaka}}, \bibinfo {author} {\bibfnamefont {M.}~\bibnamefont {Nitta}}, \
  and\ \bibinfo {author} {\bibfnamefont {R.}~\bibnamefont {Yokokura}},\ }\href
  {\doibase 10.1007/JHEP01(2021)173} {\bibfield  {journal} {\bibinfo  {journal}
  {JHEP}\ }\textbf {\bibinfo {volume} {01}},\ \bibinfo {pages} {173} (\bibinfo
  {year} {2021}{\natexlab{b}})},\ \Eprint {http://arxiv.org/abs/2009.14368}
  {arXiv:2009.14368 [hep-th]} \BibitemShut {NoStop}%
\end{thebibliography}
\end{document}